\documentstyle[fleqn,aps]{revtex}
\input psfig.sty
 
\def\c12ag{$^{12}C(\alpha,\gamma)^{16}O$}
\def\n16{$^{16}N$}
 
\begin{document}
\draft
\title{Comment on Constraints on the low-energy E1 cross
section of \c12ag from the $\beta$-delayed $\alpha$
spectrum of \n16
\footnote{Work Supported by USDOE Grant%
No. DE-FG02-94ER40870.}}
 
\author{Ralph H. France III and Moshe Gai}
 
\address{Laboratory for Nuclear Science, Dept. of Physics, UBox-46, 
University of Connecticut,\\
2152 Hillside Rd., Storrs, CT 06269-3046 \\
and \\
A.W. Wright Nuclear Structure Laboratory, Box 208124,
Dept. of Physics, Yale University, 
\\272 Whitney Ave, New Haven, CT 06520-8124}
 
\maketitle
 
\newpage
 
\begin{abstract}
 
Shortened manuscript to comply with PRC length requirements.

We dispute the alteration by Azuma {\em et al.} of the energy calibration 
of the Mainz('71) data on the beta-delayed alpha-particle emission of 
$^{16}N$ as well as the very justification of the recalibration.
We use the unaltered data to observe a pronounced disagreement between
the TRIUMF('94) and Mainz('71) data sets on both the high and low energy 
sides of the primary peak (at 2.36 MeV) of the $^{16}N$ alpha-spectrum. 
We cannot support the dismissal of the Mainz('71) spectrum 
by the TRIUMF collaboration and emphasize the need to include it in 
R-matrix fits. We discuss the need for new improved data to 
resolve this disagreement and its implication for the extracted 
p-wave astrophysical S-factor 
of the $^{12}C(\alpha,\gamma)^{16}O$ reaction.
 
\end{abstract}
 
 
A measurement of the beta-delayed alpha-particle emission of \n16
was performed at TRIUMF \cite{Bu93,Az94}, which together with an R-matrix 
analysis of these and related data, was used 
to extract the p-wave astrophysical 
S-factor of the \c12ag reaction. Such an analysis relies upon
accurate knowledge of the line-shape of the spectrum of the beta-delayed
alpha-particle emission of \n16.
In the same paper \cite{Az94} a comparison
with the Mainz('71) data is shown, as communicated
to Dr. F.C. Barker by Dr. H. W\"{a}ffler \cite{Ba96} 
and published \cite{Ha69,Ha70,Ne74}, 
and it is claimed \cite{Az94} that the Mainz('71) 
spectrum "...is difficult to fit..."
due to a broader line-shape.  Hence the Mainz('71) data have been
largely ignored by these and other authors.  In this comment we
discuss the above mentioned comparison and the
validity of the Mainz('71) data and suggest that the Mainz('71) data 
should not be ignored in future R-matrix fits due to its high statistics 
(32 million counts as compared to 1 million of the TRIUMF data).
 
Recently, we submitted for publication 
a comment on Azuma {\em et al.} \cite{Az94} 
which led to an ERRATUM \cite{Az97} explaining the reason for the 
change in the original Mainz('71) data, thus compelling us to resubmit 
so as to comment on the ERRATUM as well.  Azuma {\em et al.}
\cite{Az97} make the (implicit) claim that 
the energy calibration (10.6 keV/ch) contained
in W\"{a}ffler's communication to Barker \cite{Ba96} is wrong, and that 
the Mainz('71) spectrum can be self calibrated with high
accuracy. They use W\"{a}ffler's statement (in his letter) that
"...channel 37 corresponds to 1281 keV..." 
\cite{Ba96} and claim that the centroid 
of the $2^+$ state is accurately extracted
from the Mainz('71) spectrum \cite{Ba96}.  Using the 
known energy of the $2^+$  
state (only!) they derive a different energy dispersion (not quoted 
\cite{Az97} but most likely 
10.45 keV/ch \cite{Ba97}). This recalibration leads to a 
(very) different spectrum 
with energy shifts with "The difference 
ranges from 6.5 keV at the low end to 18 keV at the higher 
energies" \cite{Po96}.

At first we comment that it seems arbitrary that Azuma {\em et al.}
\cite{Az97} adopt a part of W\"{a}ffler's calibration ("channel 37
corresponds to 1281 keV") \cite{Ba96}, but reject the very dispersion 
(10.6 keV/ch) used to calibrate channel 37.  This illogical 
argumentation in and of itself should cast doubt on Azuma {\em et al.}
\cite{Az97}; nonetheless, in this case their recalibration \cite{Az97} 
hinges on the ability to extract the centroid
of the $2^+$ state in the Mainz('71) spectrum \cite{Ba96}.
In Fig. 1a we show the Mainz('71) data over the region of interest. 
The raw data show a very strong 
energy dependence, and in the vicinity of channel 106 one observes
a minuscule excess of counts, most likely due to a 
contribution from the $2^+$ state. 
An accurate extraction of a
centroid for this excess is very dependent on the choice of background
and requires data with extremely good statistics. In addition a reanalysis 
of the data at this time requires a knowledge 
of experimental artifacts (e.g. differential 
non-linearity of the data recording etc.) which no longer exists.

The exact energy dependence
of the background cannot be calculated {\em ab initio} as it is a convolution
of the beta-decay phase space with contributions from the broad $1^-$
state plus non-calculable background state(s). In Fig. 1a we show  
a fourth order polynomial background fit 
($\chi ^2 /\nu \ = \ 0.7$ for channels 92-100
and $\chi ^2 /\nu \ = \ 5.2$ for channels 111-114).
The background subtracted data are shown in Fig. 1b together with a fit 
to a gaussian with a centroid fixed at the 
expected energy of the $2^+$ state ($E_\alpha$ = 2.0115 MeV, 
channel 105.9). The resultant fit is not  
inconsistent with the expected shape for a $2^+$ state 
together with undulations in the background 
(more evident at the higher channels, e.g. $\chi ^2 /\nu \ = \ 5.2$ 
for channels 111-114). In addition, we emphasize that 
choices of background lead to a sizable (at least one channel) 
systematic uncertainty in the 
extracted centroid, which is different than the statistical uncertainty 
(derived from chi-square considerations).
In contrast, Azuma {\em et al.} \cite{Az97} imply an extraction 
of a centroid (with a sub-channel accuracy) 
that is approximately one channel higher than the expected location of the 
$2^+$ state \cite{Ba97}. We emphasize that this implied 
accuracy of Azuma {\em et al.} \cite{Az97} is considerably better than 
claimed by the Mainz group itself using the reliable 
$^{10}B(n,\alpha)^7Li$ calibration procedure ($\pm 10$ keV) 
\cite{Ha69,Ha70}. We conclude that the claim that the Mainz
calibration is wrong cannot be substantiated, nor can the 
recalibration procedure of Azuma {\em et al.} \cite{Az97}
be justified.

In Fig. 2 we show the ratio of the original unaltered Mainz('71) data
and the TRIUMF('94) data \cite{Az94}.
The two data sets were normalized to each
other at the highest yield point at approximately $E_{cm} \ = \ 2.36$ MeV;
we employed linear interpolations whenever necessary,
and the error bars include
the uncertainty of both data sets. This ratio deviates from
unity by as much as 30-40\%,
with a chi-square per data point of 123. Pronounced
deviations from unity are observed on both the high and low energy sides
of the highest yield point (at $E_{cm}$ = 2.36 MeV), 
in contrast to the statements of Refs. 
\cite{Az94,Az97}. The discrepancy
on the high energy side of the \n16 data negates the assertion \cite{Az94}
that it is due to partial charge collection in the Mainz
detector affecting their line-shape. This disagreement underlines 
the need for new reliable data.  In this context we refer the 
reader to our recent  
(peer reviewed and soon to be published) preliminary report \cite{Fr96}
that includes detailed figures comparing the TRIUMF data 
to new data measured by our group 
as well as a comparison with an unpublished (but extensively discussed) 
data communicated to us by the Seattle group \cite{Zh95}.  As shown 
in Ref. \cite{Fr96}, these 
new data disagree with the TRIUMF data but 
agree with the unaltered Mainz('71) singles data.
  
It is most important to evaluate the effect of the various data sets on
the extracted p-wave astrophysical S-factor of the \c12ag reaction.
This question is beyond the scope of this short
comment, but we remark that the \n16 spectrum allows for extracting
the reduced alpha-particle width of the bound $1^-$ state at 7.12 MeV,
but it can not determine a priori 
whether the interference between the bound and quasi-bound $1^-$ states
is constructive or destructive, and we doubt the validity of the
rather strong statement of Azuma {\em et al.} \cite{Az94} that their data
rule out the small S-factor solution (i.e $S_{E1}<20$ keV-b). Clearly 
a change in the line shape by as much as a factor of two at 1.4 MeV 
(the region of the interference minimum) \cite{Fr96} 
is expected to, for example, significantly
alter the f-wave contribution and thus the
extracted p-wave astrophysical S-factor of the \c12ag reaction.

We wish to acknowledge extensive communications
from Fred C. Barker and James Powell, and discussions
with James D. King.

\newpage
 
\centerline{\psfig{figure=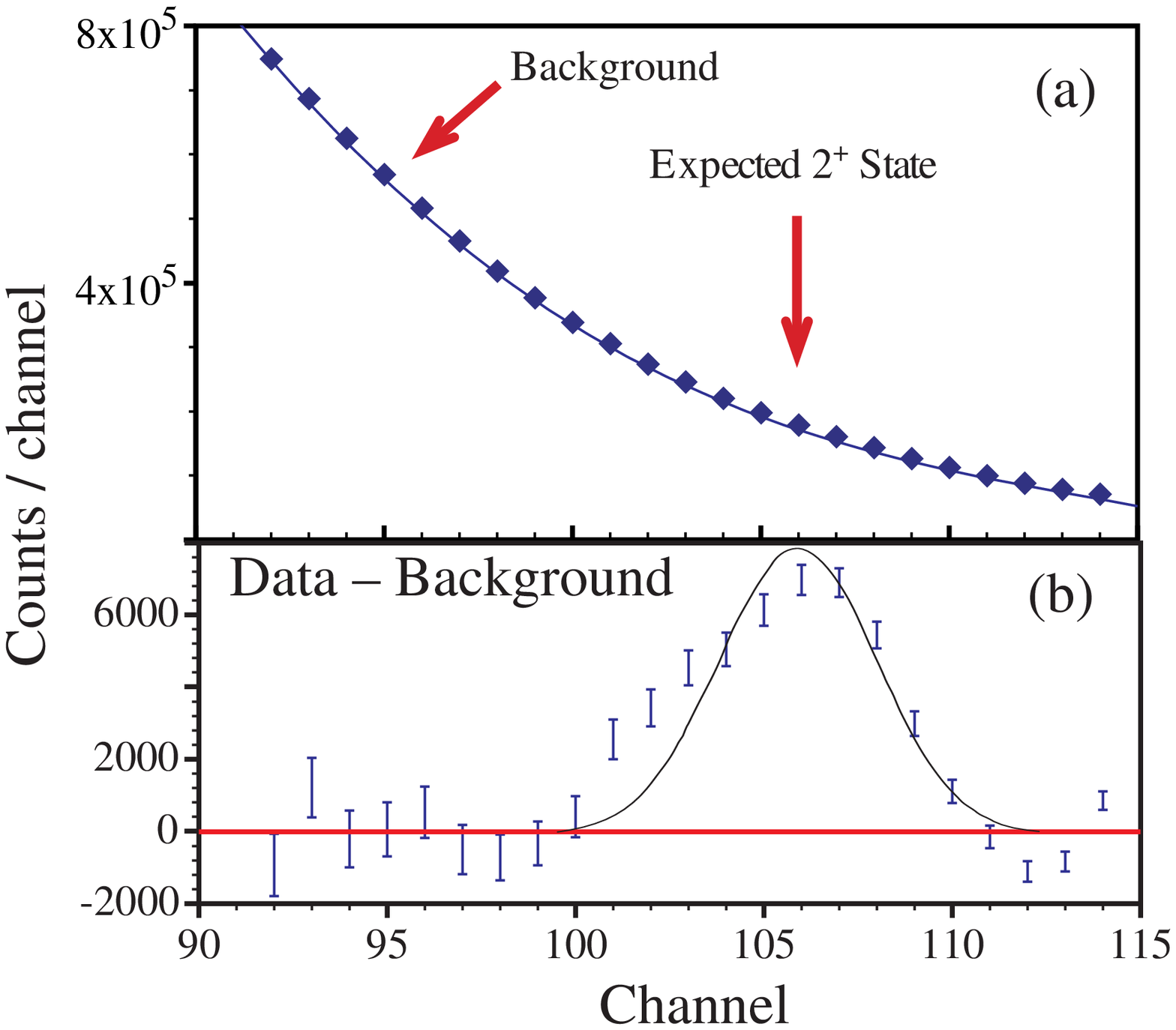,height=6.0in}}
 
\underline{Fig. 1:}  (a) The Mainz('71) data \cite{Ba96} with 
a background fit, (b) and subtracted from the data 
with a gaussian fit, as discussed in the text.
 
\newpage
 
\centerline{\psfig{figure=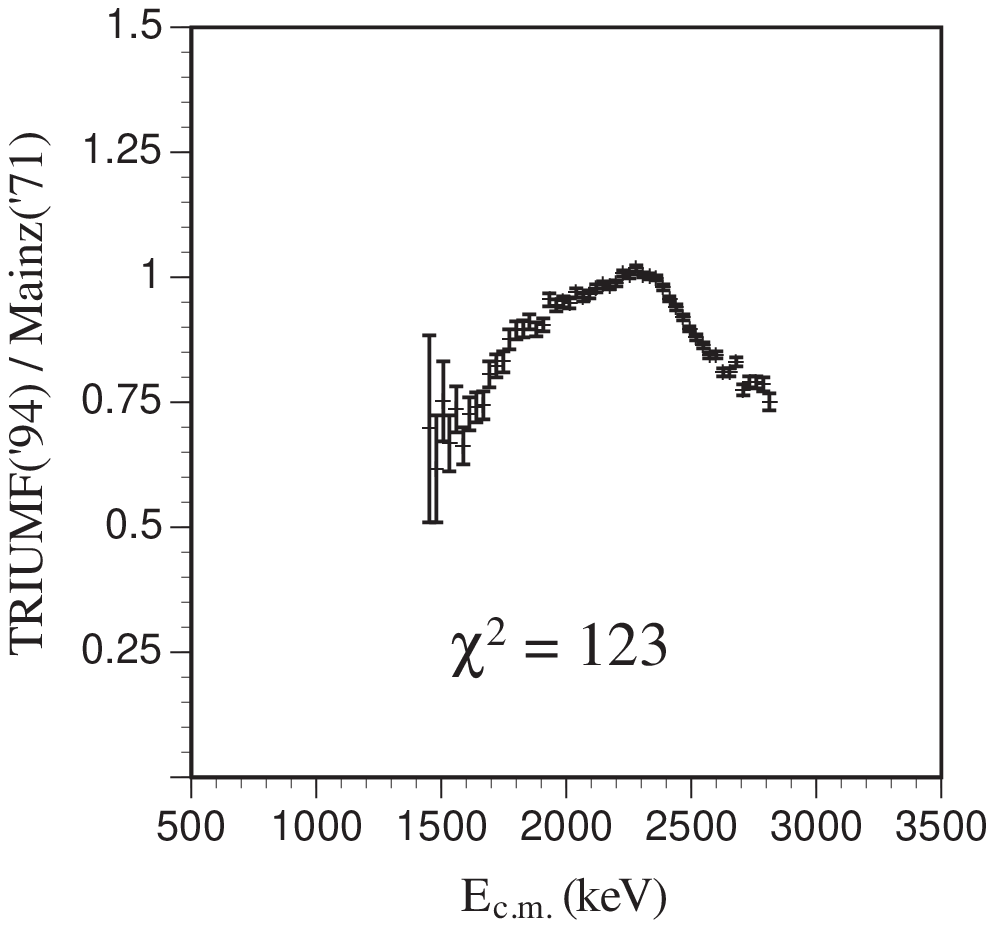,height=6.0in}}
 
\underline{Fig. 2:}  (a) Ratio of the TRIUMF('94) data \cite{Az94} to
the Mainz('71) data \cite{Ba96}, as discussed in the text.

\end{document}